\documentclass[a4paper,11pt]{article}
\pdfoutput=1
\usepackage{float}
\usepackage{graphicx}
\usepackage{dcolumn}
\usepackage{bm}
\usepackage{latexsym}
\usepackage{subfig}
\usepackage{amsmath}
\usepackage{jcappub} 

\usepackage[T1]{fontenc} 

\newcommand{\be}{\begin{equation}}
\newcommand{\ee}{\end{equation}}
\newcommand{\ba}{\begin{eqnarray}}
\newcommand{\ea}{\end{eqnarray}}

\def\({\left(}
\def\){\right)}

\title{Exploring the cosmological consequences of JLA supernova data with improved flux-averaging technique}

\author[a,1]{Shuang Wang,\note{Corresponding author.}}
\author[a]{Sixiang Wen,}
\author[a]{Miao Li}

\affiliation[a]{School of Physics and Astronomy, Sun Yat-Sen University, Zhuhai 519082, P. R. China}

\emailAdd{wangshuang@mail.sysu.edu.cn}
\emailAdd{wensx@mail2.sysu.edu.cn}
\emailAdd{limiao9@mail.sysu.edu.cn}

\abstract{In this work, we explore the cosmological consequences of the ``Joint Light-curve Analysis'' (JLA) supernova (SN) data by using an improved flux-averaging (FA) technique, in which only the type Ia supernovae (SNe Ia) at high redshift are flux-averaged.
Adopting the criterion of figure of Merit (FoM) and considering six dark energy (DE) parameterizations, we search the best FA recipe that gives the tightest DE constraints in the $(z_{cut}, \Delta z)$ plane, where $z_{cut}$ and $\Delta z$ are redshift cut-off and redshift interval of FA, respectively. Then, based on the best FA recipe obtained, we discuss the impacts of varying $z_{cut}$ and varying $\Delta z$, revisit the evolution of SN color luminosity parameter $\beta$, and study the effects of adopting different FA recipe on parameter estimation.
We find that: (1) The best FA recipe is $(z_{cut} = 0.6, \Delta z=0.06)$, which is insensitive to a specific DE parameterization. (2) Flux-averaging JLA samples at $z_{cut} \geq 0.4$ will yield tighter DE constraints than the case without using FA. (3) Using FA can significantly reduce the redshift-evolution of $\beta$. (4) The best FA recipe favors a larger fractional matter density $\Omega_{m}$. In summary, we present an alternative method of dealing with JLA data, which can reduce the systematic uncertainties of SNe Ia and give the tighter DE constraints at the same time. Our method will be useful in the use of SNe Ia data for precision cosmology.}

\begin{document}
\maketitle
\flushbottom

\section{Introduction}  \label{sec:intro}

Type Ia supernovae (SNe Ia) are a sub-category of cataclysmic variable stars
that results from the violent explosion of a white dwarf star in a binary system \citep{Hillebrandt2000}.
SNe Ia can be used as standard candle to measure the cosmic distance \citep{Riess1998,Perlmutter1999,Astier2006},
and thus have become one of the most powerful tools to probe the nature of dark energy (DE)
\citep{Tsujikawa2006,Frieman2008,Kamionkowski2009,Cai2010,LiLiWangWang2011,LiLiWangWang2013,Weinberg2013,WangWangLi2016}.
In the recent decade, several high quality supernova (SN) data-sets were released,
such as ``Union'' \citep{Kowalski2008}, ``Constitution'' \citep{Hicken2009a,Hicken2009b}, ``SDSS'' \citep{Kessler2009},
``Union2'' \citep{Amanullah2010}, ``SNLS3'' \citep{Conley2011} and ``Union2.1'' \citep{Suzuki2012}.
In a latest paper, the ``Joint Light-curve Analysis'' (JLA) data \citep{Betoule2014}, which consisting of 740 SNe Ia, was released.
This SN sample has been widely studied in the literature.

As the rapid growth of the number of SNe discovered in the astronomical observation,
the control of the systematic uncertainties of SNe Ia have drawn more and more attention in recent years.
For examples, the studies on various SNe Ia data-sets, such SNLS3 \citep{WangWang2013}, Union2.1 \citep{Mohlabeng2013},
Pan-STARRS1 \citep{Scolnic14} and JLA \citep{Shariff2015,li2016}, all indicated that SN color luminosity parameter $\beta$
should evolve along with redshift $z$ at high confidence level (CL);
it has also been proved that the evolution of $\beta$ has significant effects on the parameter estimation of various cosmological models \citep{WLZ2014,WWGZ2014,WWZ2014,WGHZ2015}.
Besides, it was found that the intrinsic scatter $\sigma_{\rm int}$ has the hint of redshift-dependence
that will affect the results of cosmology fits \citep{Marrinerl11}.
In addition, another factor that causes SN systematic uncertainties is the choice of light-curve fitters (LCF) \citep{Kessler2009}:
for instance, adopting ``MLCS2k2'' \citep{Jha07} and ``SALT2'' \citep{Guy07} LCF
will lead to completely different fitting results \citep{Pigozzo11,Bengochea14};
in contrast, using ``SALT2'' and ``SiFTO'' \citep{Conley2011} LCF will yield the consistent fitting results \citep{hu2015}.

Instead of analysing the factors of SN systematic uncertainties one by one,
an alternative is to deal with these factors in an unified framework.
In 2000, Wang proposed a data analysis technique, called flux-averaging (FA) \citep{Wang2000},
to reduce the systematic uncertainties caused by the weak lensing of SNe Ia.
The key idea is that due to flux conservation in gravitational lensing,
the average flux of a large number of SNe Ia at the same redshift should be unity;
thus averaging the observed flux from a large number of SNe Ia at the same redshift
can recover the unlensed brightness of the SNe Ia at that redshift \citep{Wang2004}.
In a series of research works, Wang and the collaborators have applied the FA technique
to give the cosmological constraints on DE \citep{wangprl2004,Wang2006,Wang2007,YunWang2008},
and have found that using FA can also reduce the bias in distance estimate induced by some other systematic effects \citep{WangTegmark05,YunWang2009,Wang12CM}.

The original FA method is to average the observed flux of all the SN samples.
For this case, FA only relates to one quantity: the redshift interval $\Delta z$.
In \citep{WangWang2013}, one of the present author and Wang proposed an improved version of FA,
in which only the SN data at high-redshift are flux-averaged.
For this new FA method, another quantity, redshift cut-off $z_{cut}$, is introduced.
For the SN samples at $z < z_{cut}$, the $\chi^2$ is computed by using the usual ``magnitude statistics'';
for the SN samples at $z \geq z_{cut}$, the $\chi^2$ is computed by using the ``flux statistics''.
In a recent work, Wang and Dai applied this improved FA technique to explore the JLA SN sample,
and found that it can give tighter constraints on DE \citep{Wang2015}.
However, some important factors are not taken into account in \citep{Wang2015}.
For example, only one kind of FA recipe, $(z_{cut} = 0.5, \Delta z=0.04)$, is used in \citep{Wang2015}.
This choice of $(z_{cut}, \Delta z)$ may not be the best FA recipe that can give the tightest DE constraints.
In addition, only one kind of DE parameterization,
i.e. the Chevallier-Polarski-Linder (CPL) model \citep{Chevallier2001, Linder2003}, is used in \citep{Wang2015}.
So the conclusions of \citep{Wang2015} may depends on the specific model.
Moreover, only the fitting results of the CPL model are briefly discussed in \citep{Wang2015},
while the impacts of adopting FA on the systematic uncertainties of JLA sample are not discussed.

The aim of our work is to present a systematic and comprehensive investigation
on the cosmological consequences of the JLA data-set by using the improved FA method.
Compared with previous literature, the novelty of our work are as follows:
(1) Quite different from the previous studies,
we scan the whole $(z_{cut}, \Delta z)$ plane to search the best FA recipe,
by adopting the criterion of the dark energy task force (DETF) figure of Merit (FoM) \citep{Albrecht}.
(2) To ensure that our results are insensitive to a specific DE model,
six kinds of DE parameterizations, including the CPL model, the Wetterich model \citep{Wetterich2004},
the Jassal-Bagla-Padmanabhan (JBP) model \citep{Jassal2005a, Jassal2005b},
the Barbosa-Alcaniz (BA) model \citep{Barboza2008}, the Wang model ~\citep{YunWang2008}
and the Felice-Nesseris-Tsujikawa (FNT) model~\citep{fnt2012} are taken into account.
(3) Based on the best FA recipe obtained in this work, we discuss the impacts of varying $z_{cut}$ and varying $\Delta z$, revisit the evolution of SN color luminosity parameter $\beta$, and study the effects of adopting different FA recipe on parameter estimation.

In addition to the JLA SN data,
we also use the galaxy clustering (GC) measurements of Hubble parameter $H(z)$ and angular diameter distance $D_A(z)$ \citep{Wang2015}
extracted from the Sloan Digital Sky Survey (SDSS) at $z = 0.35$ (SDSS DR7 \citep{CW12}) and $z = 0.57$ (BOSS DR11 \citep{Anderson14}),
as well as the cosmic microwave background (CMB) distance priors
$(l_a, R, \Omega_{b}h^2)$ derived from the 2015 Planck data \citep{Planck201514}.

We describe our method in Section~\ref{sec:method}, present our results in Section~\ref{sec:results}, and summarize in Section~\ref{sec:conclusion}.

\section{Methodology}
\label{sec:method}

In this section, we briefly introduce the six DE parameterization models,
and then describe the observational data used in the present work.

\subsection{Theoretical Model}
\label{subsec:theoretical model}
In a flat universe,
\footnote{The assumption of flatness is motivated by the inflation scenario.
For a detailed discussion of the effects of spatial curvature, see \citep{Clarkson07}}
the comoving distance $r(z)$ to an object is given by
\be \label{eq:rz}
r(z)= cH_0^{-1}\int_0^z\frac{dz'}{E(z')},
\ee
where $c$ is the speed of light, and $H_0$ is the Hubble constant.
$E(z)$ is the reduced Hubble parameter, which satisfies
\be\label{E_z}
E(z)=\sqrt{\Omega_{r}(1+z)^{4}+\Omega_{m}(1+z)^{3}+\Omega_{de}X(z)},
\ee
where $\Omega_{r}$, $\Omega_{m}$ and $\Omega_{de}$
are the present fractional densities of radiation, matter and DE, respectively.
Per Ref. \citep{WangyunWangshuang2013}, we take $\Omega_{r}=\Omega_{m}/(1+z_{\rm eq})$,
where $z_{\rm eq}=2.5\times10^4\Omega_{m}h^2(T_{\rm cmb}/2.7\,{\rm K})^{-4}$ \cite{EisenHu98},
$T_{\rm cmb}=2.7255\,{\rm K}$, and $h$ is the Hubble constant.
Note that $\Omega_{r}+\Omega_{m}+\Omega_{de} = 1$, so $\Omega_{de}$ is not a model parameter.
In addition, $X(z)$ is the DE density function, which satisfies
\be\label{Density function.e.}
X(z)={\rm exp}\Big[3\int_{0}^{z}dz^{\prime}\frac{1+w(z^{\prime})}{1+z^{\prime}}\Big].
\ee
Here $w$ is the equation of state (EoS) of DE, which is the key element to determine the properties of DE.
For the simplest $\Lambda$-cold-dark-matter ($\Lambda$CDM) model, $X=1$.

To study the impacts of different $w(z)$, here we consider six popular parametrization models:
\begin{itemize}
\item
 The CPL model. It has a dynamical EoS $w(z) = w_{0} + w_{a}\frac{z}{1+z}$, then
\be
X(z)=(1+z)^{3(1+w_0+w_a)}e^{\frac{-3w_az}{1+z}}.
\ee
\item
 The Wetterich model \citep{Wetterich2004}. It has a dynamical EoS $w(z) = \frac{w_{0}}{[1+w_aln(1+z)]^2}$, then
\be
 X(z)=(1+z)^{3+\frac{3w_0}{1+w_aln(1+z)}}.
\ee
\item
The JBP model \citep{Jassal2005a, Jassal2005b}. It has a dynamical EoS $w(z) = w_{0} + w_{a}\frac{z}{(1+z)^2}$, then
\be
X(z)=(1+z)^{3(1+w_0)}e^{\frac{3w_az^2}{2(1+z)^2}}.
\ee
\item
The BA model \citep{Barboza2008}. It has a dynamical EoS $w(z) = w_{0} + w_{a}\frac{z(1+z)}{1+z^2}$, then
\be
X(z)=(1+z)^{3(1+w_0)}(1+z^2)^{\frac{3w_a}{2}}.
\ee
\item
The Wang model~\citep{YunWang2008} has a dynamical EoS $w(z) = w_{0}\frac{1-2z}{1+z} + w_{a}\frac{z}{(1+z)^2}$, so
\be
X(z)=(1+z)^{3(1-2w_0+3w_a)}e^{\frac{9(w_0-w_a)z}{1+z}}.
\ee
\item
 The FNT model~\citep{fnt2012}. It has a dynamical EoS $w(z) = w_{a} + (w_0-w_{a})\frac{a[1-(a/a_t)^{1/\tau}]}{1-a_t^{-1/\tau}}$, then
\be
X(z)=(1+z)^{3(1+w_a)}e^{3(w_0-w_a)\frac{1+(1-a_t^{-1/\tau})\tau+((\tau((1+z)a_t)^{-1/\tau}-1)-1)/(1+z)}{(1+\tau)(1-a_t^{-1/\tau})}},
\ee
 where $a_t$ and $\tau$ are two new parameters.
\end{itemize}
For each model, the parameter $w_0$ represent the current value of EoS,
and the parameter $w_a$ reflect the evolution of EoS.

In 2006, the DETF  team proposed a quantity, called FoM ~\citep{Albrecht},
to quantitatively assess the ability of constraining DE of a experiment project.
$FoM$ is defined as the reciprocal of the error ellipse area
enclosing the $95\%$ confidence limit in the $w_0-w_a$ plane of the CPL model.
It satisfies
\be \label{FoM}
FoM = \frac{1}{6.17 \pi [detCov(w_0, w_a)]^{1/2}},
\ee
where $Cov(w_0, w_a)$ is covariance matrix of $w_0$ and $w_a$.
It is clear that a larger $FoM$ indicates a better accuracy.
In this work, we extend the application scope of $FoM$ to the cases of other DE parametrization.
Moreover, as mentioned above, based on the criterion of $FoM$,
we will search the best FA recipe that can give the tightest DE constraints.

\subsection{Observational Data}
\label{subsec:observational data}

In this section, firstly, we introduce how to calculate the $\chi^2$ of JLA data using the usual ``magnitude statistics''.
Then, we introduce how to calculate the $\chi^2$ of JLA data using the ``flux statistics'',
as well as the details of the improved FA method.
At last, we briefly introduce other observational data used in this work.

Theoretically, the distance modulus $\mbox{\bf $\mu$}_{th}$ in a flat universe can be written as
\be
  \mbox{\bf $\mu$}_{th} = 5 \log_{10}\bigg[\frac{d_L(z_{hel},z_{cmb})}{Mpc}\bigg] + 25,
\ee
where $z_{cmb}$ and $z_{hel}$ are the CMB restframe and heliocentric redshifts of SN.
The luminosity distance ${d}_L$ is given by
\be
  {d}_L(z_{hel},z_{cmb}) = (1+z_{hel}) r(z_{cmb}),
\ee
where $r(z)$ has been given in Eq. \ref{eq:rz}.

The observation of distance modulus $\mbox{\bf $\mu$}_{obs}$ is given by a
empirical linear relation:
\be
  \mbox{\bf $\mu$}_{obs}= m_{B}^{\star} - M_B + \alpha \times {\bf x}_1
  -\beta \times {\cal C},
\ee
where $m_B^{\star}$ is the observed peak magnitude in the rest-frame
\text{of the} $B$ band,
$x_1$ describes the time stretching of light-curve, ${\cal C}$ describes the
supernova color at maximum brightness and $M_B$ is the absolute B-band magnitude,
which depends on the host galaxy properties \citep{Schlafly11,Johansson13}.
Notice that $M_B$ is related to the host stellar mass $M_{stellar}$ by a simple step function \citep{Betoule2014}
\begin{equation}
  \label{eq:mabs}
    M_B = \left\lbrace
   \begin{array}{ll}
    M^1_B &\quad \text{if}\quad  M_\text{stellar} < 10^{10}~M_{\odot}\,,\\
    M^2_B &\quad \text{otherwise.}
    \end{array}
    \right.
\end{equation}
Here $M_{\odot}$ is the mass of sun.

The $\chi^2$ of JLA data can be calculated as
\be  \label{eq:chi2_SN}
  \chi^2_{SN} = \Delta \mbox{\bf $\mu$}^T \cdot \mbox{\bf Cov}^{-1} \cdot \Delta\mbox{\bf $\mu$},
\ee
where $\Delta \mbox{\bf $\mu$}\equiv \mbox{\bf $\mu$}_{obs}-\mbox{\bf $\mu$}_{th}$
is the data vector and $\mbox{\bf Cov}$ is the total covariance matrix, which is given
by
\be
\mbox{\bf Cov}=\mbox{\bf D}_{\rm stat}+\mbox{\bf C}_{\rm stat}
+\mbox{\bf C}_{\rm sys}.
\ee
Here $\mbox{\bf D}_{\rm stat}$ is the diagonal part of the statistical
uncertainty, which is given by \citep{Betoule2014},
\begin{eqnarray}
\mbox{\bf D}_{\rm stat,ii}&=&\left[\frac{5}{z_i \ln 10}\right]^2 \sigma^2_{z,i}+
  \sigma^2_{\rm int} +\sigma^2_{\rm lensing} + \sigma^2_{m_B,i} +\alpha^2 \sigma^2_{x_1,i}+\beta^2 \sigma^2_{{\cal C},i}\nonumber\\
&&+ 2 \alpha C_{m_B x_1,i} - 2 \beta C_{m_B {\cal C},i}  -2\alpha\beta C_{x_1 {\cal C},i},
\end{eqnarray}
where the first three terms account for the uncertainty in redshift due to peculiar velocities,
the intrinsic variation in SN magnitude and the variation of magnitudes caused by gravitational lensing.
$\sigma^2_{m_B,i}$, $\sigma^2_{x_1,i}$, and $\sigma^2_{{\cal C},i}$
denote the uncertainties of $m_B$, $x_1$ and ${\cal C}$ for the $i$-th SN.
In addition, $C_{m_B x_1,i}$, $C_{m_B {\cal C},i}$ and $C_{x_1 {\cal C},i}$
are the covariances between $m_B$, $x_1$ and ${\cal C}$ for the $i$-th SN.
Moreover, $\mbox{\bf C}_{\rm stat}$ and $\mbox{\bf C}_{\rm sys}$
are the statistical and the systematic covariance matrices, given by
\begin{equation}
\mbox{\bf C}_{\rm stat}+\mbox{\bf C}_{\rm sys}=V_0+\alpha^2 V_a + \beta^2 V_b +
2 \alpha V_{0a} -2 \beta V_{0b} - 2 \alpha\beta V_{ab},
\end{equation}
where $V_0$, $V_{a}$, $V_{b}$, $V_{0a}$, $V_{0b}$ and $V_{ab}$ are matrices given by the JLA group.

As pointed out in \citep{Betoule2014}, in the process of calculating $\chi^2_{SN}$,
the absolute B-band magnitude $M_B$ is marginalized.
So in this work, we follow the procedure of \citep{Betoule2014}, and do not treat $M_B$ as free parameter.
We refer the reader to Ref. \citep{Betoule2014} for the details of calculating $\chi^2_{SN}$.

Now, let us turn to the FA of JLA data.
FA is a very useful technique to reduce the bias in distance estimate
induced by various systematic effects \citep{Wang2004,WangTegmark05,YunWang2009}.
The original FA method divide the whole redshift region of SNe Ia into a lot of bins,
where the redshift interval of each bin is $\Delta z$.
Then, for the JLA SN data in each bin,
the steps of FA are as follows \citep{Wang12CM}:

(1) Convert the distance modulus of SNe Ia into ``fluxes'',
\be
\label{eq:flux}
F(z_l) \equiv 10^{-(\mu_{obs}(z_l)-25)/2.5} =
\left( \frac{d_L^{\rm {obs}}(z_l)} {\mbox{Mpc}} \right)^{-2}.
\ee
Here $z_l$ represent the CMB restframe redshift of SN and $\mu_{obs}(z_l)$ is the observation of distance modulus for the $l$-th SN at $z_l$.

(2) For a set of free model parameters $\{ {\bf s} \}$ whose values are not fixed a priori,
obtain ``absolute luminosities'' \{${\cal L}(z_l)$\} by removing the redshift dependence of the ``fluxes'', i.e.,
\be
\label{eq:lum}
{\cal L}(z_l) \equiv d_L^2(z_l |{\bf s})\,F(z_l).
\ee

(3) Flux-average the ``absolute luminosities'' \{${\cal L}^i_l$\}
in each redshift bin $i$ to obtain $\left\{\overline{\cal L}^i\right\}$:
\be
 \overline{\cal L}^i = \frac{1}{N_i}
 \sum_{l=1}^{N_i} {\cal L}^i_l(z^{(i)}_l),
 \hskip 1cm
 \overline{z_i} = \frac{1}{N_i}
 \sum_{l=1}^{N_i} z^{(i)}_l.
\ee

(4) Place $\overline{\cal L}^i$ at the mean redshift $\overline{z}_i$ of
the $i$-th redshift bin, now the binned flux is
\be
\overline{F}(\overline{z}_i) = \overline{\cal L}^i /
d_L^2(\overline{z}_i|\mbox{\bf s}).
\ee
with the corresponding flux-averaged distance modulus:
\be
\overline\mu^{obs}(\overline{z}_i) =-2.5\log_{10}\overline{F}(\overline{z}_i)+25.
\ee

(5) Compute the covariance matrix of $\overline{\mu}(\overline{z}_i)$
and $\overline{\mu}(\overline{z}_j)$:
\be
\mbox{Cov}\left[\overline{\mu}(\overline{z}_i),\overline{\mu}(\overline{z}_j)\right]
=\frac{1}{N_i N_j \overline{\cal L}^i \overline{\cal L}^j}
 \sum_{l=1}^{N_i} \sum_{m=1}^{N_j} {\cal L}(z_l^{(i)})
{\cal L}(z_m^{(j)}) \langle \Delta \mu_0^{\rm { obs}}(z_l^{(i)})\Delta
\mu_0^{\rm {obs}}(z_m^{(j)})
\rangle
\ee
where $\langle \Delta \mu_0^{\rm { obs}}(z_l^{(i)})\Delta \mu_0^{\rm { obs}}(z_m^{(j)})\rangle $
is the covariance of the measured distance moduli of the $l$-th SN Ia
in the $i$-th redshift bin, and the $m$-th SN Ia in the $j$-th
redshift bin. ${\cal L}(z)$ is defined by Eqs.(\ref{eq:flux}) and (\ref{eq:lum}).

(6) For the flux-averaged data, $\left\{\overline{\mu}(\overline{z}_i)\right\}$,
compute
\be
\label{eq:chi2_SN_fluxavg}
\chi^2_{SN} = \sum_{ij} \Delta\overline{\mu}(\overline{z}_i) \,
\mbox{Cov}^{-1}\left[\overline{\mu}(\overline{z}_i),\overline{\mu}(\overline{z}_j)
\right] \,\Delta\overline{\mu}(\overline{z}_j)
\ee
where
\be
\Delta\overline{\mu}(\overline{z}_i) \equiv
\overline{\mu}(\overline{z}_i) - \mu^p(\overline{z}_i|\mbox{\bf s}),
\ee
and
\be
\overline\mu^p(\overline{z}_i) =-2.5\log_{10} F^p(\overline{z}_i)+25.
\ee
with $F^p(\overline{z}_i|\mbox{\bf s})=
\left( d_L(z|\mbox{\bf s}) /\mbox{Mpc} \right)^{-2}$.

As mentioned above, the improved FA method  \citep{WangWang2013} introduce a new quantity: the redshift cut-off $z_{cut}$.
For the SN samples at $z < z_{cut}$, the $\chi^2$ is computed by using the usual ``magnitude statistics'' (i.e., Eq. \ref{eq:chi2_SN});
for the SN samples at $z \geq z_{cut}$, the $\chi^2$ is computed by using the ``flux statistics'' (i.e., Eq. \ref{eq:chi2_SN_fluxavg}).
In \citep{Wang2015}, Wang and Dai applied this improved FA method to explore the JLA data-set,
and found that it can give tighter constraints on DE.
However, only one kind of FA recipe, $(z_{cut} = 0.5, \Delta z=0.04)$, is considered in \citep{Wang2015}.
This choice of $z_{cut}$ and $\Delta z$ may not be the best FA recipe that can give the tightest DE constraints.
In order to find out the best FA recipe, by adopting the criterion of DETF $FoM$,
we scan the whole $(z_{cut}, \Delta z)$ plane in this work.
Specifically, we set that $z_{cut} = 0.1i$, $i=0,1,2,...,8$ and $\Delta z = 0.01j$, $j=1,2,3,...,10$.

In addition to the JLA SN data, we also use the GC and the CMB data to improve the constraint results of DE.

For the GC data, we use the measurements of $H(z)r_s(z_d)/c$ (where $H(z)$ is the Hubble parameter)
and $D_A(z)/r_s(z_d)$ (where $D_A(z)=\frac{r(z)}{1+z}$ is the angular diameter distance)\citep{Wang2015}
extracted from the SDSS DR7 at $z = 0.35$ \citep{CW12} and the BOSS DR11 at $z = 0.57$ \citep{Anderson14}.
The $r_s(z_d)$ is the sound horizon at the drag epoch, which can be fitted by \citep{WangWang2013}
\be~\label{eq:rs}
r_s(z) = cH_0^{-1}\int_{0}^{a}\frac{da^{\prime}}{\sqrt{3(1+\overline{R_b}a^\prime){a^\prime}^4E^2(z^\prime)}},
\ee
where $\overline{R_b}=31500\Omega_{b}h^2(T_{cmb}/2.7K)^{-4}$, and $\Omega_{b}$ is the present fractional density of baryon.
The redshift of the drag epoch $z_d$ can be well approximated by \cite{EisenHu98}
\begin{equation}
z_d  =
 \frac{1291(\Omega_mh^2)^{0.251}}{1+0.659(\Omega_mh^2)^{0.828}}
\left[1+b_1(\Omega_bh^2)^{b2}\right],
\label{eq:zd}
\end{equation}
where
\begin{eqnarray}
  b_1 &= &0.313(\Omega_mh^2)^{-0.419}\left[1+0.607(\Omega_mh^2)^{0.674}\right],\\
  b_2 &= &0.238(\Omega_mh^2)^{0.223}.
\end{eqnarray}
Thus, the $\chi^2$ function for the GC data can be expressed as
\be
\label{eq:chi2bao}
\chi^2_{GC}=\sum_{i}\chi^2_{GCi}=\sum_{i}\Delta p_i \left[ {\rm C}^{-1}_{GC}(p_i,p_j)\right]
\Delta p_j,
\hskip .5cm
\Delta p_i= p_i - p_i^{data},
\ee
where $p_1=H(z_{GCi})r_s(z_d)/c$ and $p_2=D_A(z_{GCi}/r_s(z_d)$,
with $i=1,2$.

For the CMB data, we use the distance priors $(l_a, R, \Omega_{b}h^2)$ derived from the 2015 Planck data \citep{Planck201514}.
\footnote{In addition to \citep{Planck201514},
there are some other CMB distance priors data, e.g. see Refs. \citep{WangyunWangshuang2013,Huang2015,Wang2015}.}
The acoustic scale $l_A$ is defined as
\begin{equation}
\label{ladefeq} l_A\equiv \pi r(z_*)/r_s(z_*),
\end{equation}
where $r(z_*)$ is the comoving distance, and $r_s(z_*)$ is the comoving sound
horizon at $z_*$. We use the
fitting function of $z_*$ proposed by \cite{Hu:1995en}:
\begin{equation}
\label{zstareq} z_*=1048[1+0.00124(\Omega_b
h^2)^{-0.738}][1+g_1(\Omega_m h^2)^{g_2}],
\end{equation}
where
\begin{equation}
g_1=\frac{0.0783(\Omega_b h^2)^{-0.238}}{1+39.5(\Omega_b
h^2)^{0.763}},\quad g_2=\frac{0.560}{1+21.1(\Omega_b h^2)^{1.81}}.
\end{equation}
In addition, the shift parameter $R$ is given by ~\citep{WangMukherjee2007}:
\be
R \equiv \sqrt{\Omega_{m} H_0^2} \,r(z_*)/c.
\ee
Thus, the $\chi^2$ function for the CMB data can be expressed as
\be
\label{eq:chi2CMB}
\chi^2_{CMB}=\Delta p_i \left[ \mbox{Cov}^{-1}_{CMB}(p_i,p_j)\right]
\Delta p_j,
\hskip .2cm
\Delta p_i= p_i - p_i^{data},
\ee
where $p_1=R(z_*)$, $p_2=l_a(z_*)$, and $p_3=\Omega_{b}h^2$.

We perform a MCMC likelihood analysis \citep{Lewis2002} to obtain $O(10^6)$ samples for each set of results presented in this paper.
In this work we choose $(\alpha, \beta, \Omega_{m}, \Omega_{b} h^2, h, w_0, w_a)$ as a set of free parameters.
The prior range of each free parameter is uniform for all the DE models. These prior ranges are listed in Table~\ref{tab:1}.

\begin{table*}
\centering
\caption{The prior ranges of the free parameters. These prior ranges are uniform for all the models considered in this work.}
\label{tab:1}
\centering

\begin{tabular}{ccccccccccc}
\hline\hline 
&&\\
Parameter  & $\alpha$ & $\beta$ & $\Omega_{b} h^2$ & $h$& $\Omega_{m}$ & $w_0$ & $w_a$ \\
&&\\ \hline
&&\\
Prior Range                      & $[-4,5]$
                                & $[0,10]$
                                & $[0,1]$
                                & $[0,2]$
                                & $[0,0.8]$
                                & $[-3,2]$
                                & $[-15,15]$ $\ ^a$ \\ 
&&\\
\hline
\end{tabular}
\leftline{\noindent$\ ^a$ For the CPL model and the Wetterich model, the prior range of $w_a$ are $[-10,10]$ and $[0,15]$.}
\end{table*}

\section{Result}
\label{sec:results}

This section include four parts.
Firstly, by scanning the whole $(z_{cut}, \Delta z)$ plane, we find out the best FA recipe that can yield the largest value of DETF $FoM$.
Secondly, based on the CPL model and the best FA recipe, we discuss the impacts of varying $z_{cut}$ and varying $\Delta z$.
Thirdly, by combining FA with redshift tomography technique, we revisit the evolution of $\beta$ for the JLA data.
fourthly, by using the JLA data alone, we study the impacts of adopting different FA recipe on parameter estimation.

\subsection{Searching The Best FA Recipe}
\label{subsec:de parametrizations}

\begin{figure*}
  \centering
  \includegraphics[height=9cm]{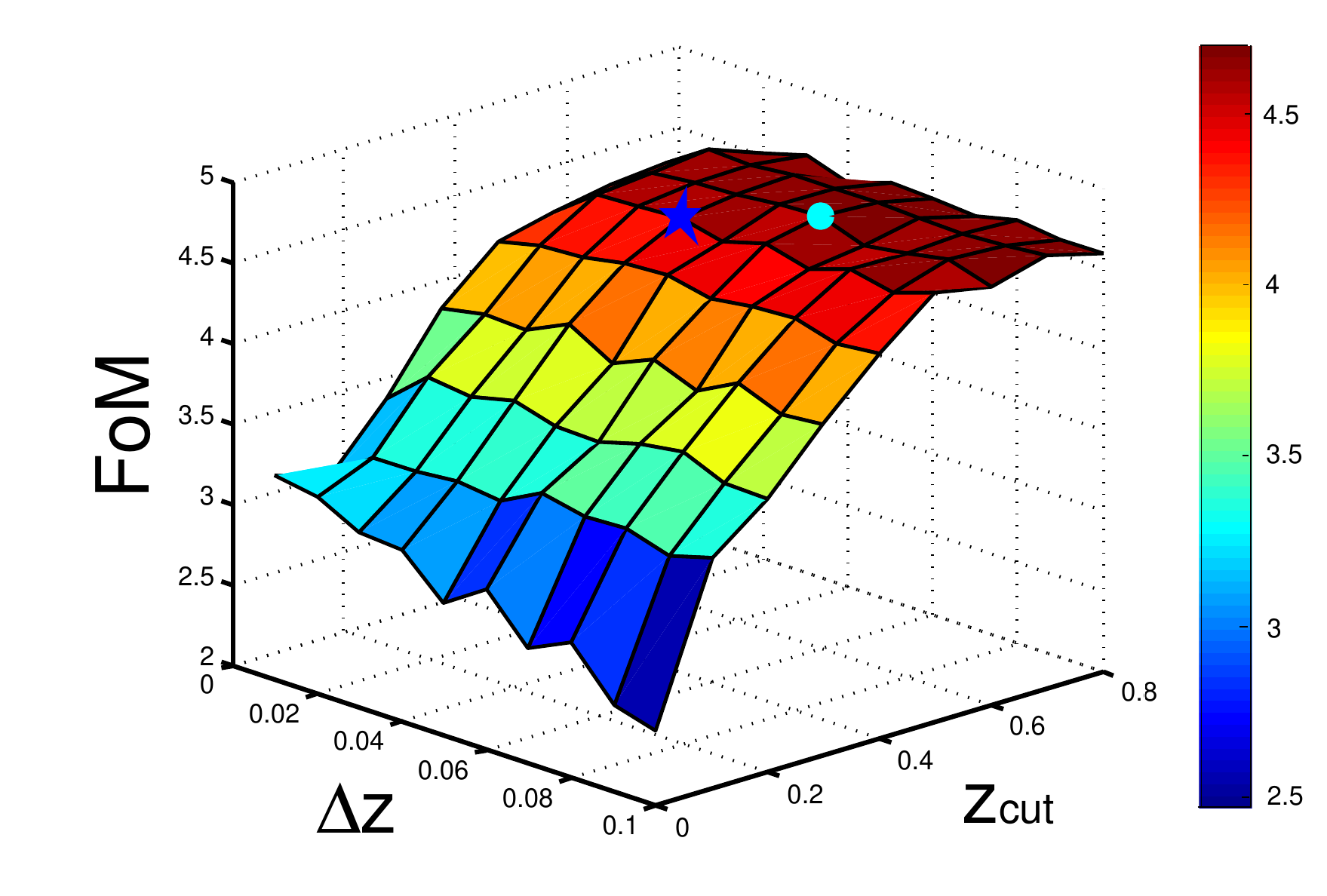}
  \caption{The 3D graph that show the results of $FoM$ given by different choices of $(z_{cut}, \Delta z)$,
  for the CPL model.
  Here the combined SN+GC+CMB data are used in the analysis.
  Note that $z_{cut} = 0.1i$, $i=0,1,2,...,8$; while $\Delta z = 0.01j$, $j=1,2,3,...,10$.
  The cyan round dot denotes the best FA recipe $(z_{cut} = 0.6, \Delta z=0.06)$, which can give a largest $FoM = 4.6965$.
  The blue star represents the recipe of WD16 FA \citep{Wang2015} $(z_{cut} = 0.5, \Delta z=0.04)$, which give a $FoM = 4.6083$.}

\label{fig:1}
\end{figure*}

To search the best FA recipe, we scan the whole $(z_{cut}, \Delta z)$ plane.
For each choice of $(z_{cut}, \Delta z)$,
we perform a MCMC analysis for the CPL model, and then calculate the corresponding value of $FoM$.
In Fig. \ref{fig:1}, we plot a 3D graph,
which show the results of $FoM$ given by different choices of $(z_{cut}, \Delta z)$.
It should be mentioned that the combined SN+GC+CMB data are used in the analysis.
From this figure we can see that, the original FA method (i.e., $z_{cut}=0$) always give a small $FoM$;
in contrast, the improved FA recipe with a $(z_{cut}$ at high redshift will yield a larger $FoM$.
This means that the improved FA recipe can give better DE constraints than the original FA recipe.
Moreover, we find out the best FA recipe $(z_{cut} = 0.6, \Delta z=0.06)$ (the blue star),
which can give a largest $FoM = 4.6965$.
As a comparison, in \citep{Wang2015},
Wang and Dai adopted the FA recipe $(z_{cut} = 0.5, \Delta z=0.04)$ (the blue round dot), which give a $FoM = 4.6083$.
therefore, we can conclude that our FA recipe can give tighter DE constraints than the recipe of \citep{Wang2015}.
This shows the importance of finding out the best FA recipe.

\begin{table*}
\tiny
  \caption{ Cosmology-fits results for the CPL, the Wetterich, the JBP, the BA and the Wang models, given by different FA recipe.
  Both the best-fit result of various parameters and the corresponding results of $FoM$ are listed in this table.
  ``No FA'' denote the case without using FA,
  ``Original FA'' represent the FA recipe $(z_{cut}=0,\Delta z=0.06)$,
  ``Best FA'' correspond to the FA recipe $(z_{cut}=0.6,\Delta z=0.06)$.
  The combined SN+GC+CMB data are used in the analysis.}

\label{tab:2}

\begin{tabular}{cp{0.45cm}p{0.55cm}p{0.35cm}cp{0.45cm}p{0.55cm}p{0.35cm}cp{0.45cm}p{0.55cm}p{0.35cm}cp{0.45cm}p{0.55cm}p{0.35cm}cp{0.45cm}p{0.55cm}p{0.35cm}}
\hline\hline &\multicolumn{3}{c}{CPL}&&\multicolumn{3}{c}{Wetterich}&&\multicolumn{3}{c}{JBP}&&\multicolumn{3}{c}{BA}&&\multicolumn{3}{c}{ Wang} \\
  \cline{2-4}\cline{6-8}\cline{10-12}\cline{14-16}\cline{18-20}
Parm  & No FA & Original FA & Best FA && No FA & Original FA & Best FA & & No FA & Original FA & Best FA & &No FA & Original FA & Best FA & & No FA & Original FA & Best FA \\ \hline
  $\alpha$         & $0.140$ %
                   & $0.080$  %
                   & $0.140$ &
                   & $0.141$
                   & $0.091$
                   & $0.141$&
                   & $0.141$
                   & $0.090$
                   & $0.140$&
                   & $0.141$
                   & $0.082$
                   & $0.139$&
                   & $0.140$
                   & $0.075$
                   & $0.139$\\

$\beta$          & $3.109$  %
                   & $3.088$   %
                   & $3.258$ &
                   & $3.081$
                   & $3.058$
                   & $3.231$&                    %
                   & $3.096$
                   & $3.019$
                   & $3.252$&
                   & $3.098$
                   & $3.067$
                   & $3.220$&
                   & $3.094$
                   & $3.117$
                   & $3.232$\\

$h$          & $0.684$   %
                   & $0.683$   %
                   & $0.675$&
                   & $0.684$
                   & $0.687$
                   & $0.672$&
                   & $0.683$
                   & $0.687$
                   & $0.679$&
                   & $0.681$
                   & $0.685$
                   & $0.670$&
                   & $0.682$
                   & $0.686$
                   & $0.671$\\

$\Omega_m$      & $0.299$  %
                   & $0.298$  %
                   & $0.305$ &  %
                   & $0.299$
                   & $0.297$
                   & $0.308$&
                   & $0.299$
                   & $0.297$
                   & $0.301$&
                   & $0.301$
                   & $0.298$
                   & $0.310$&
                   & $0.301$
                   & $0.296$
                   & $0.309$\\

$w_0$              & $-0.996$  %
                   & $-1.002$  %
                   & $-0.976$ &  %
                   & $-0.988$
                   & $-1.000$
                   & $-0.963$&
                   & $-0.996$
                   & $-1.052$
                   & $-1.070$&
                   & $-0.984$
                   & $-0.993$
                   & $-0.943$&
                   & $-0.974$
                   & $-1.010$
                   & $-0.940$\\

$w_a$              & $0.026$  %
                   & $0.078$  %
                   & $0.119$ & %
                   & $0.0039$
                   & $0.0045$
                   & $0.0384$&
                   & $0.094$
                   & $0.323$
                   & $0.684$&
                   & $0.011$
                   & $-0.006$
                   & $0.028$&
                   & $-0.986$
                   & $-0.986$
                   & $-0.938$\\

 $FoM$             & $4.320$
                   & $3.001$
                   & $4.696$          &
                   & $20.188$
                   & $14.931$
                   & $21.762$&
                   & $1.829$
                   & $1.179$
                   & $1.843$&
                   & $8.831$
                   & $6.079$
                   & $9.057$&
                   & $13.138$
                   & $8.905$
                   & $14.044$\\ 

\hline
\end{tabular}
\end{table*}


Let us discuss this topic with more details.
Table~\ref{tab:2} show the cosmology-fits results for the CPL, the Wetterich, the JBP, the BA and the Wang models, given by different FA recipe.
Since these is no enough space to list the results of all the six models in one table,
here we do not show the specific results of the FNT model, which lead to the same conclusions with the other five models.
Both the best-fit result of various parameters and the corresponding results of $FoM$ are listed in this table.
``No FA'' denote the case without using FA,
``Original FA'' represent the FA recipe $(z_{cut}=0,\Delta z=0.06)$,
``Best FA'' correspond to the FA recipe $(z_{cut}=0.6,\Delta z=0.06)$.
From this table we can see that, for all the DE models,
adopting the ``Best FA'' recipe will yield a larger $\beta$ and a larger $\Omega_m$,
compared with the cases of ``No FA'' and ``Original FA''.
Moreover, among these five FA recipes,
the ``Original FA'' recipe always give the smallest $FoM$,
while the ``Best FA'' recipe always give the largest $FoM$.
This means that the ``Best FA'' recipe can yield tightest DE constraints.
Since this result holds true for all the six DE parameterizations,
our conclusion is insensitive to the specific DE model considered in the background.

\subsection{The Impacts of Varying $z_{cut}$ And Varying $\Delta z$}

\begin{figure*}
  \centering
  \includegraphics[height=5.1cm]{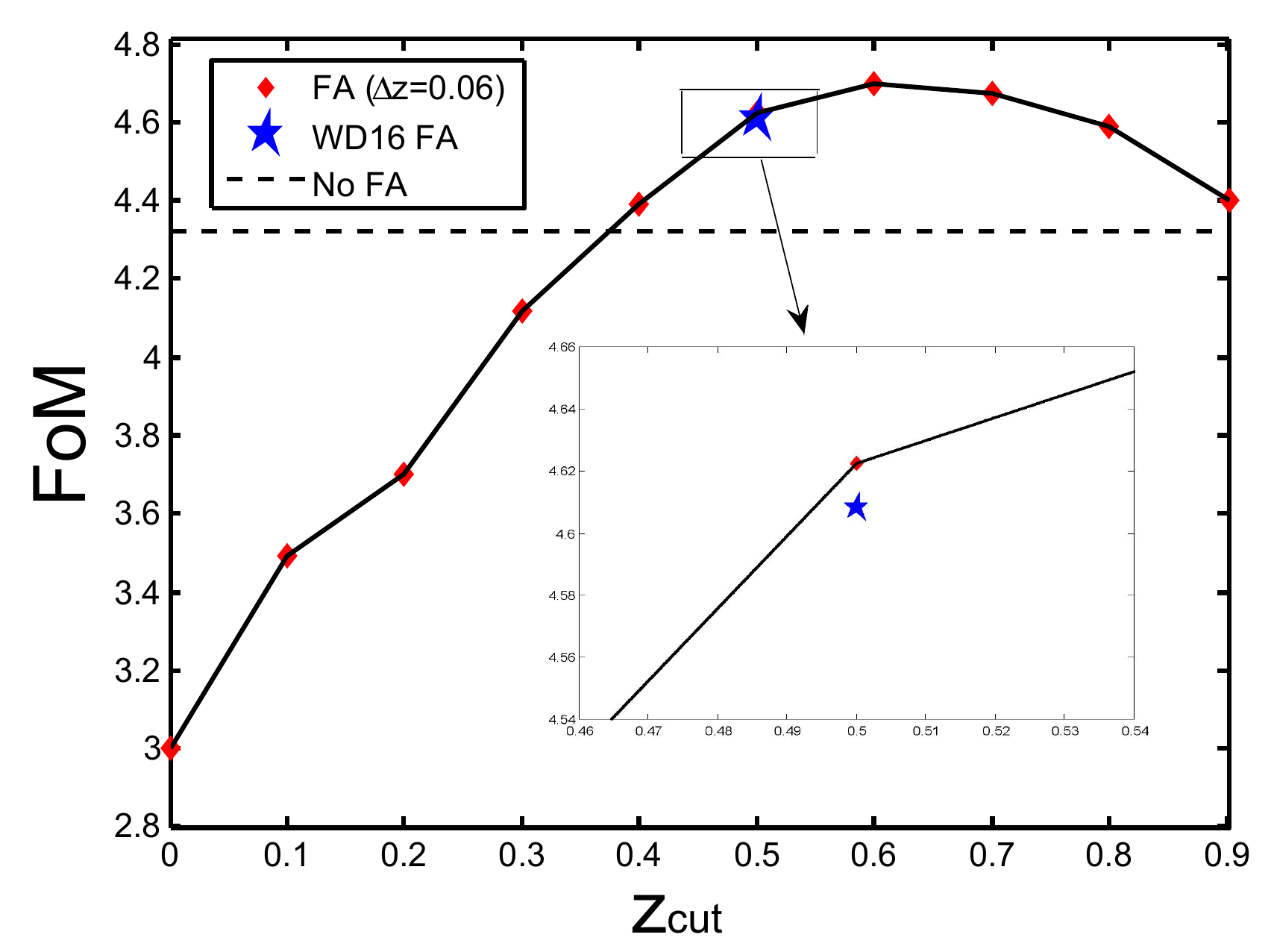}
  \includegraphics[height=5.1cm]{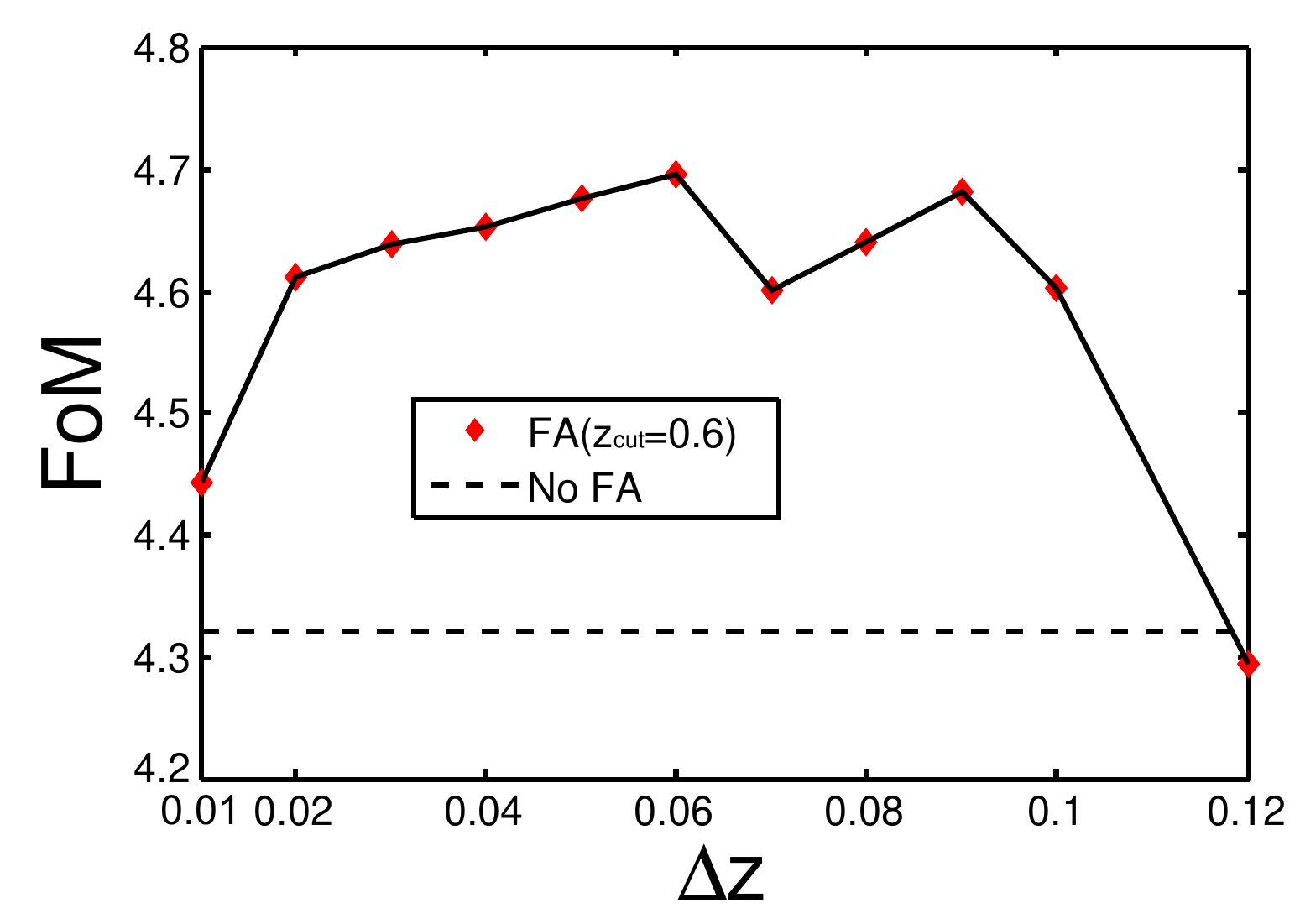}
  \caption{The results of $FoM$ given by different $z_{cut}$ (left panel) and different $\Delta z$ (right panel), for the CPL model.
  Here the combined SN+GC+CMB data are used in the analysis.
  For the left panel, $\Delta z = 0.06$ is always fixed; for the right panel, $z_{cut} = 0.6$ is always fixed.
  For comparison, we also show the results of $FoM$ for the case without using FA (black dashed line)
  and the case using WD16 FA recipe (blue star).}
\label{fig:2}
\end{figure*}

Based on the CPL model and the best FA recipe, we further investigate the effects of varying $z_{cut}$ and varying $\Delta z$.
In Fig. \ref{fig:2}, we give results of $FoM$ given by different $z_{cut}$ (left panel) and different $\Delta z$ (right panel).
Note that the combined SN+GC+CMB data are used in the analysis.
To make a comparison, we also give the results of $FoM$ for the case without using FA (dashed line)
and for the case of WD16 FA recipe (red star).
From the left panel of Fig. \ref{fig:2}, we find that varying $z_{cut}$ will significantly affect the result of $FoM$:
if $z_{cut}<0.4$, then using FA will yield a smaller $FoM$ than the result of ``NO FA'' case;
if $z_{cut}\geq0.4$, then using FA will yield a larger $FoM$ than the result of ``NO FA'' case.
In other words, flux-averaging JLA samples at $z_{cut} \geq 0.4$ will yield tighter DE constraints than the case without using FA.
In \citep{Wang2015}, Wang and Dai argued that flux-averaging JLA samples at $z_{cut} \geq 0.5$ will yield tighter constraints.
Since only one kind of FA recipe (i.e., a point $[z_{cut},\Delta z]=[0.5,0.04]$) was considered in the WD16 paper,
we can conclude that our conclusion is more accurate.
Moreover, from the right panel of Fig. \ref{fig:2} we find that,
once $z_{cut} = 0.6$ is fixed, adopting FA (except for the case of $\Delta z = 0.12$) will always yield a larger $FoM$ than the result of ``NO FA'' case.
This means that, compared with the case of varying $z_{cut}$,
the effects of varying $\Delta z$ are much smaller.
In addition, it is clear that the value of $FoM$ decrease for $z_{cut} >0.8$ and $\Delta z >0.1$.

\subsection{The Revisit of The Redshift-Evolution of $\beta$}

In a previous paper \citep{li2016}, we studied the evolution of $\beta$ for the JLA data by using the redshift tomography (RT) technique.
The basic idea of RT is to divide the SN data into different redshift bins, assuming that $\beta$ is a piecewise constant.
Then one can constrain $\Lambda$CDM model and check the consistency of cosmology-fit results in each bin.
It was found that, for the JLA data-set, $\beta$ has a significant trend of decreasing at high redshift.
It must be stressed that only the usual ``magnitude statistics'' was used in \citep{li2016},
while the FA of SNe Ia was not taken into account.
It will be interesting to study the effects of adopting FA technique on $\beta$'s evolution.

\begin{figure*}
  \centering
  \includegraphics[height=5.7cm]{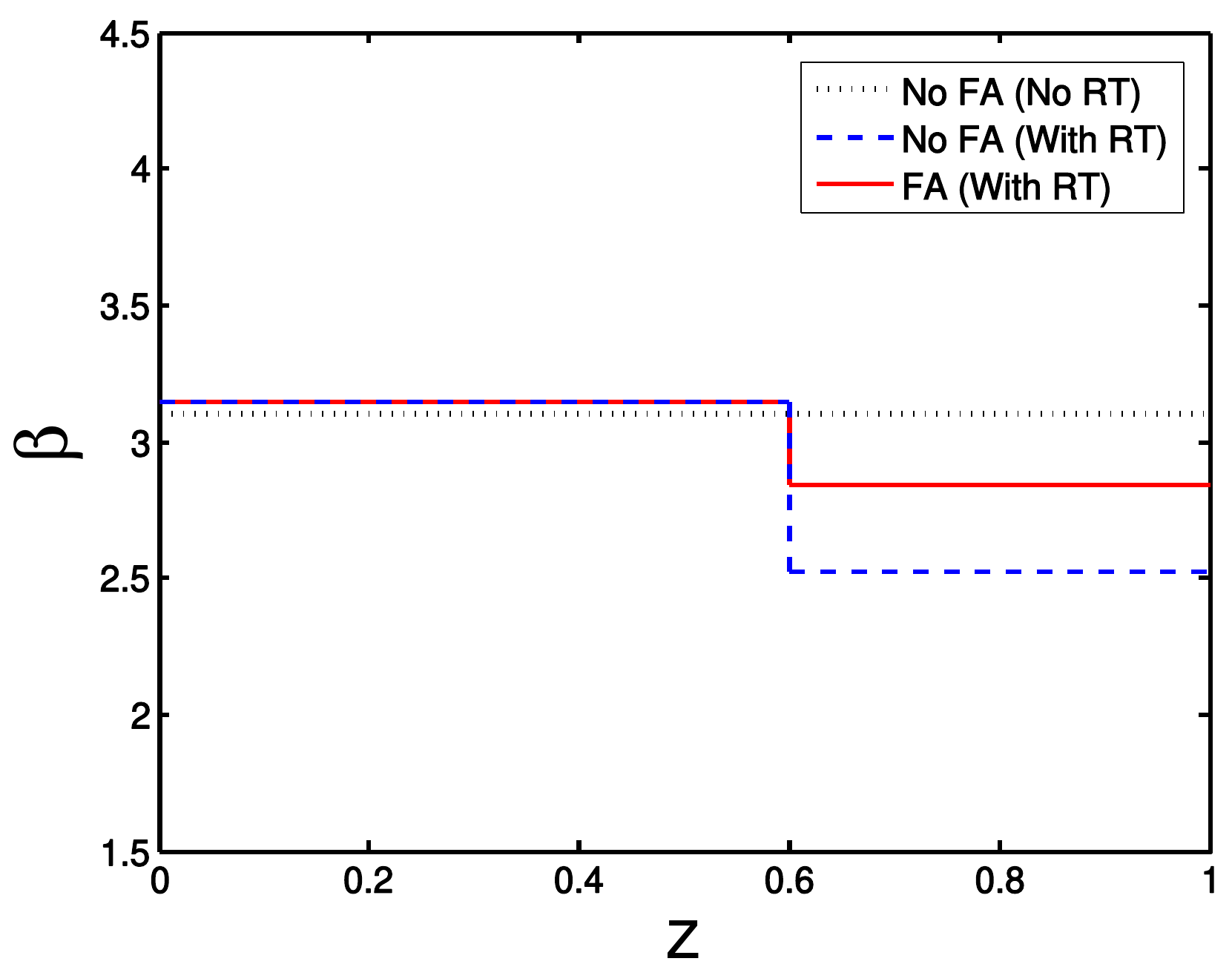}
  \includegraphics[height=5.7cm]{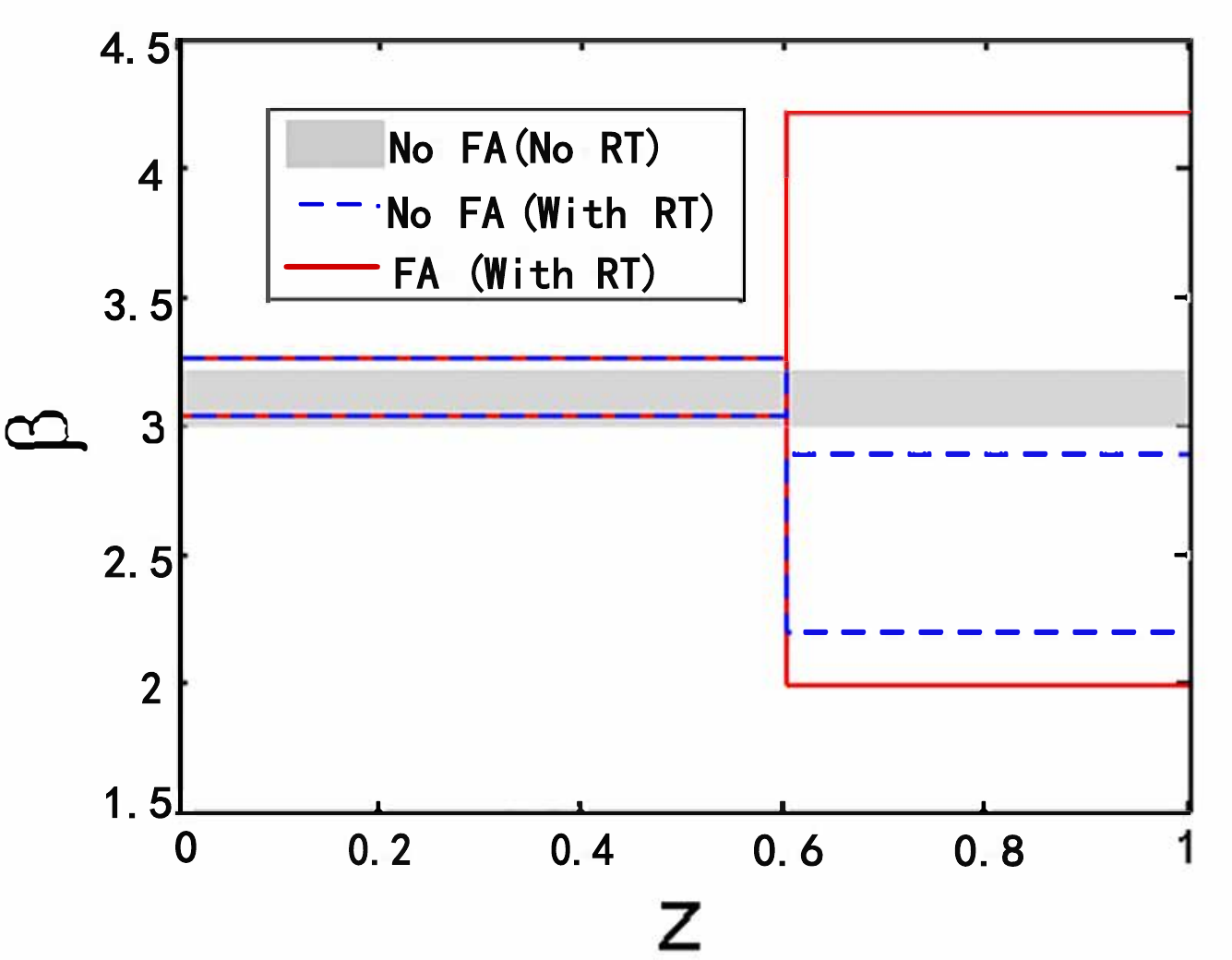}
  \caption{The best-fit results (left panel) and the $1\sigma$ regions (right panel) of $\beta$
  at redshift region $[0, 1]$ given by different SN  analysis technique.
  ``No FA (No RT)'' denotes the case without using FA and redshift tomography.
  ``No FA (With RT)'' represents the case only using redshift tomography with $z_{cut} = 0.6$.
  ``FA (With RT)'' corresponds to the case using best FA and redshift tomography with $z_{cut} = 0.6$.
  For the left panel, the balck dotted line, the red dashed line and the red solid line
  correspond to the best-fit results of $\beta$ of ``No FA (No RT)'', ``No FA (With RT)'' and ``FA (With RT)'', respectively.
  For the right panel, the grey region, the region inside blue dashed lines and the region inside red solid lines
  correspond to the $1\sigma$ regions of $\beta$ of ``No FA (No RT)'', ``No FA (With RT)'' and ``FA (With RT)'', respectively.
  Note that we adopt the $\Lambda$CDM model; besides, only the JLA SN data are used in the analysis.}
\label{fig:3}
\end{figure*}

We revisit the evolution of $\beta$ for the JLA sample by combining FA with RT technique.
For the RT, we divide the SN data into two redshift bins with a discontinuity point $z_{cut} = 0.6$;
for the FA, we adopt the FA recipe $(z_{cut} = 0.6, \Delta z=0.01)$, which has more data points at high redshift than the best FA recipe.
In Fig. ~\ref{fig:3}, we plot the best-fit results (left panel) and the $1\sigma$ regions (right panel) of $\beta$
given by different SN analysis technique.
``No FA (No RT)'' denotes the case without using FA and redshift tomography.
``No FA (With RT)'' represents the case only using redshift tomography with $z_{cut} = 0.6$.
``FA (With RT)'' corresponds to the case using best FA and redshift tomography with $z_{cut} = 0.6$.
Here we study the $\Lambda$CDM model; besides, only the JLA SN data are used in the analysis.
Note that the results of $\beta$ at low redshift give by ``No FA (With RT)'' and ``FA (With RT)'' are the same,
because the improve FA technique only flux-average SNe Ia at high redshift.
From the left panel of Fig. \ref{fig:3} we find that,
only using RT will yield a significant decrease for the best-fit result of $\beta$ at $z_{cut} \geq 0.6$,
which is consistent with result of \citep{li2016}.
In contrast, adopting FA will yield a larger best-fit value of $\beta$ at $z_{cut} \geq 0.6$,
which is much closer to the result of the ``No FA (No RT)'' case.
From the right panel of Fig. \ref{fig:3} we see that,
if only using RT, the $1\sigma$ region of $\beta$ at $z_{cut} \geq 0.6$
will significantly deviate from the $1\sigma$ region at $z_{cut} < 0.6$,
as well as the ``No FA (With RT)'' case.
In contrast, after adopting FA,
the $1\sigma$ region of $\beta$ at $z_{cut} \geq 0.6$
will be consistent with the $1\sigma$ region at $z_{cut} < 0.6$,
as well as the ``No FA (With RT)'' case.
These results show that using FA can significantly reduce the redshift-evolution of $\beta$.

\subsection{The Impacts of Adopting Different FA Recipe on Parameter Estimation}

At last, let us discuss the impacts of adopting different FA recipe on parameter estimation.
For simplicity, here we just consider the standard cosmological model: the $\Lambda$CDM model.
In addition, in this subsection, we make use of the JLA SN data alone to perform the MCMC analysis.

\begin{figure*}
  \centering
  \includegraphics[height=7cm]{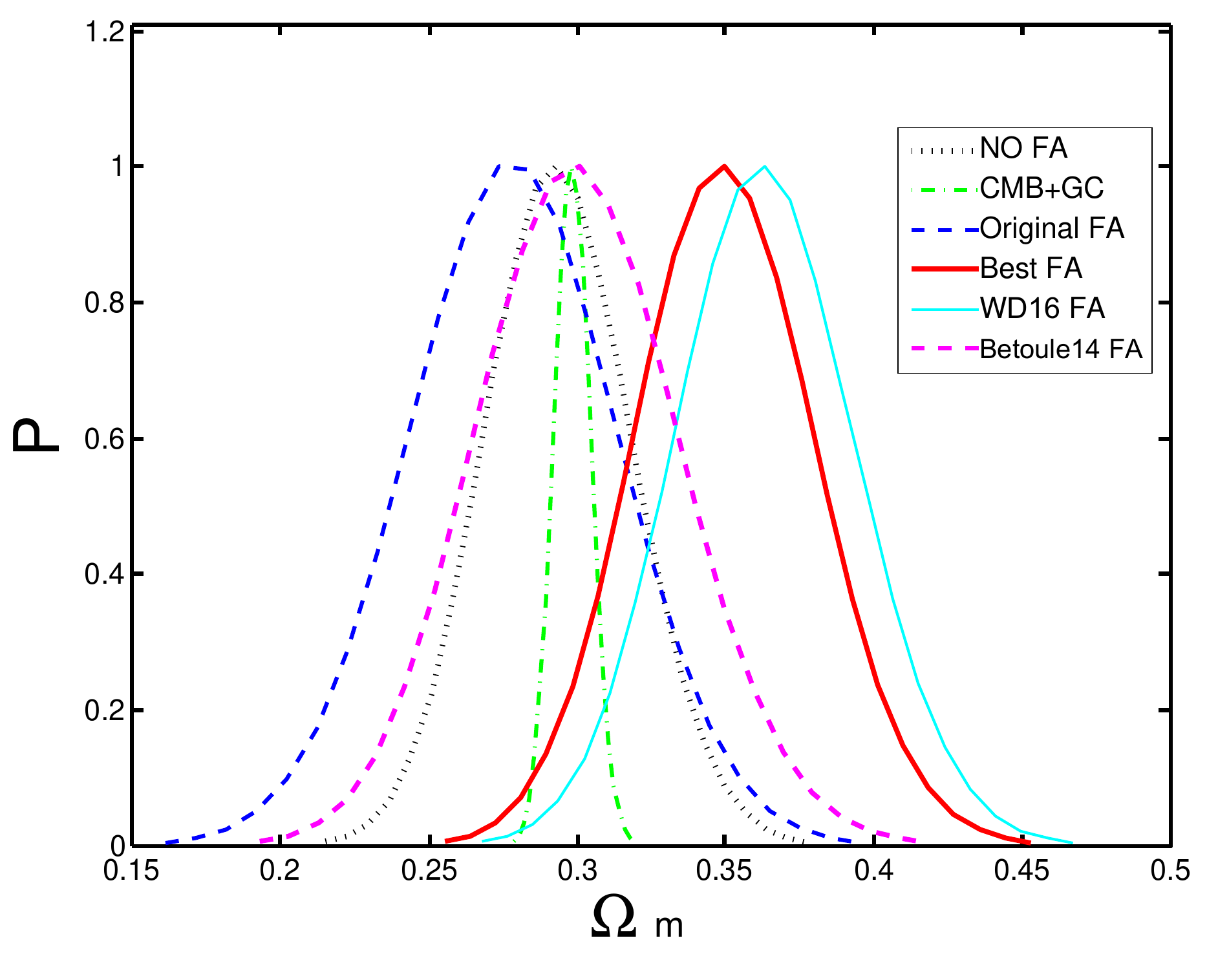}
  \caption{The 1D marginalized probability distributions of $\Omega_{m}$ given by different FA recipe, for the $\Lambda$CDM model.
  ``No FA'' (black dotted line) denotes the case without using FA,
  ``Original FA'' (blue dashed line) represents the FA recipe $(z_{cut}=0,\Delta z=0.06)$,
  ``Best FA'' (red solid line) corresponds to the FA recipe $(z_{cut}=0.6,\Delta z=0.06)$.
  ``WD 16 FA'' (cyan solid line) represents the FA recipe $(z_{cut}=0.5,\Delta z=0.04)$ from \citep{Wang2015},
  ``Betoule 14 FA'' (magenta dashed line) corresponds to the FA recipe with Betoule's binning method (see the Appendix E of \citep{Betoule2014}).
  The corresponding result given by the CMB+GC data (green dashed-dotted line) is also shown for comparison.
  Only the JLA data are used in the analysis.}
\label{fig:4}
\end{figure*}

In Fig. ~\ref{fig:4}, we plot the 1D marginalized probability distributions of $\Omega_{m}$ given by different FA recipe, for the $\Lambda$CDM model.
Note that ``No FA'' (black dotted line) denotes the case without using FA,
``Original FA'' (blue dashed line) represents the FA recipe $(z_{cut}=0,\Delta z=0.06)$,
``Best FA'' (red solid line) corresponds to the FA recipe $(z_{cut}=0.6,\Delta z=0.06)$.
It can be seen that, the original FA recipe will give a smaller $\Omega_m$,
while the best FA recipe will give a larger $\Omega_m$.
The best-fit values of $\Omega_m$ given by ``No FA'', ``Original FA'' and ``Best FA'' are 0.2929, 0.2829 and 0.3478, respectively.
\footnote{ The shift in $\Omega_m$ will disappear when the combined SN, GC and CMB data are used to constrain the $\Lambda$CDM model,
which is consistent with the results of Table~\ref{tab:2}.
This is because the CMB data have very powerful ability in constraining parameter $\Omega_m$.}
It is very interesting to note that the enlarge of $\Omega_m$ had also been found in \citep{li2016},
in which the impacts of $\beta$'s evolution on parameter estimation are taken into account.
In other word, both adopting the best FA method and considering the evolution of $\beta$ will yield a larger $\Omega_m$.
As a comparison, we also apply the FA to the ``Betoule14'' binning recipe (see the Appendix E of \citep{Betoule2014}),
which divides the SN samples at the redshift region $0.01<z<1.3$ into 30 bins evenly.
The 1D $\Omega_{m}$ distributions given by this FA recipe (magenta dashed line),
as well as the results given by the ``WD16'' FA recipe (cyan solid line) and the CMB+GC data (green dashed-dotted line),
are also plotted in Fig. ~\ref{fig:4}.
It can be seen that, there is no obvious shift in $\Omega_m$ between the ``Betoule14 FA'' and the ``No FA'' recipe;
in contrast, using the ``WD16 FA'' recipe will yield the largest $\Omega_{m}$.
Therefore, the shift in $\Omega_m$ may due to the specific redshift binning recipe.


\begin{figure*}
  \centering
  \includegraphics[height=5.7cm]{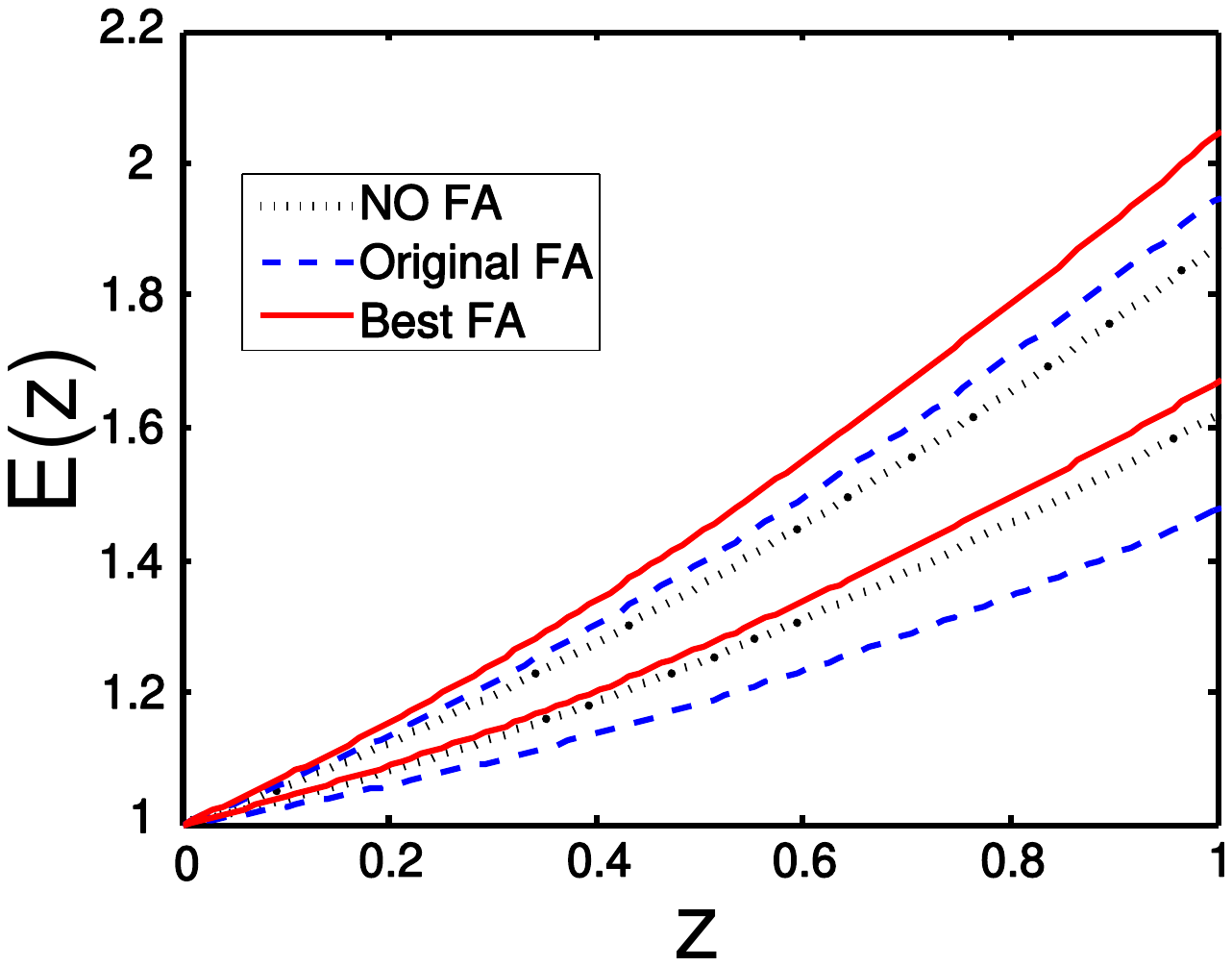}
  \includegraphics[height=5.7cm]{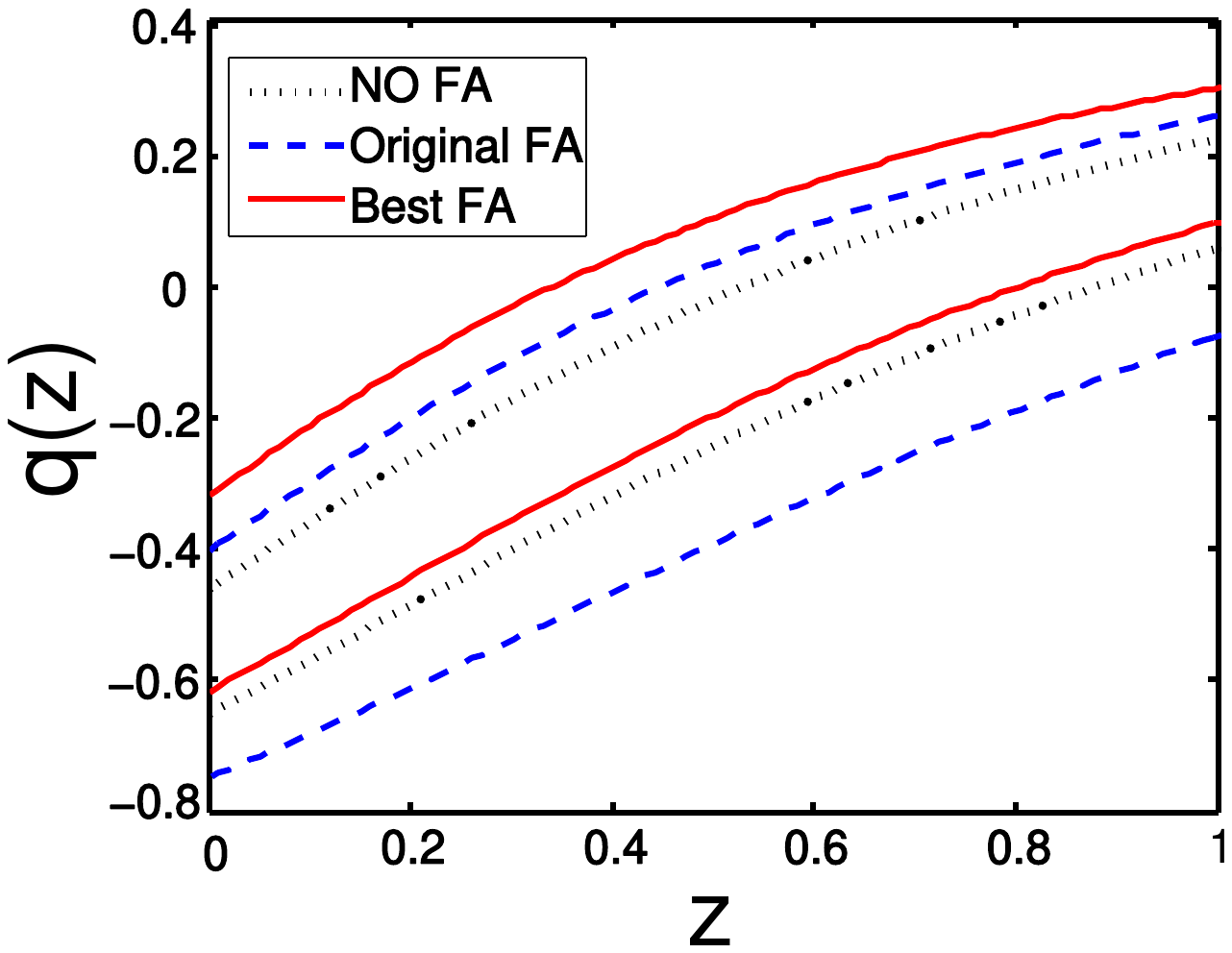}
  \caption{$1\sigma$ regions of $E(z)$ (left panel) and $q(z)$ (right panel) at redshift region $[0, 1]$
  given by different FA recipe, for the $\Lambda$CDM model.
  ``No FA'' (the region inside black dotted lines) denotes the case without using FA,
  ``Original FA'' (the region inside blue dashed lines) represents the FA recipe $(z_{cut}=0,\Delta z=0.06)$,
  ``Best FA'' (the region inside red solid lines) corresponds to the FA recipe $(z_{cut}=0.6,\Delta z=0.06)$.
  Only the JLA data are used in the analysis.
  }
\label{fig:5}
\end{figure*}

In Fig. ~\ref{fig:5}, we plot the $1\sigma$ regions
of reduced Hubble parameter $E(z)$ (left panel) and deceleration parameter $q(z) \equiv -\frac{\ddot{a}}{aH^2}$ (right panel)
at redshift region $[0, 1]$ given by different FA recipe, for the $\Lambda$CDM model.
The left panel of Fig. \ref{fig:5} shows that adopting the original FA recipe will yield a smaller $E(z)$ at high redshift,
while adopting the best FA recipe will yield a larger $E(z)$ at high redshift.
The right panel of Fig. \ref{fig:5} shows that making use of the original FA recipe will yield a smaller $q(z)$,
while adopting the best FA recipe will yield a larger $q(z)$.
In other words, the best FA recipe favors a universe with a smaller acceleration.
This result is consistent with the result of Fig. ~\ref{fig:4},
because a larger $\Omega_m$ (i.e., more matter and less DE) will correspond to a weaker effect of anti-gravity.
It should be mentioned that the conclusion of this subsection only holds true for the case of a flat $\Lambda$CDM model with SN only data.

\section{CONCLUSION AND DISCUSSION}
\label{sec:conclusion}

The control of the systematic uncertainties of SNe Ia has become a fundamental topic of precision cosmology.
Flux-averaging, which is proposed in \citep{Wang2000},
has been proved to be a very useful technique to reduce the systematic uncertainties of SNe Ia.
Note that the original FA technique will lead to a significant decrease of SN data points,
and thus will yield larger error bars for various model parameter.
To solve this problem, an improved version of FA was proposed,
in which only the SN data at high-redshift are flux-averaged \citep{WangWang2013}.
This new method relates to two quantities: $z_{cut}$ and $\Delta z$.
In previous studies, both $z_{cut}$ and $\Delta z$ are set as a specific value.
For example, in \citep{Wang2015}, Wang and Dai only considered the case of $(z_{cut} = 0.5, \Delta z=0.04)$.
However, as shown in the current work,
different choice of $(z_{cut}, \Delta z)$ will significantly change the results of $FoM$,
and arbitrarily choosing a FA recipe can not give the tightest DE constraints.

In this work, we have presented a systematic and comprehensive investigation on the cosmological consequences of the JLA SN sample
by using the improved FA technique.
Quite different from the previous studies, we have scanned the whole $(z_{cut}, \Delta z)$ plane to search the best FA recipe;
for each choice of $(z_{cut}, \Delta z)$, we have performed a MCMC analysis for the CPL parameterization,
and have calculated the corresponding value of $FoM$.
To ensure that our results do not depend on a specific DE parameterization,
we have also considered the cases of the Wetterich, the JBP, the BA, the WANG and the FNT model.
Then, based on the best FA recipe obtained, we have discussed the impacts of varying $z_{cut}$ and varying $\Delta z$.
Next, combining FA with RT technique, we have revisited the evolution of $\beta$.
Finally, using the JLA data alone, we have studied the impacts of adopting different FA recipe on parameter estimation.

Our conclusions are as follows:
\begin{itemize}

\item
(1) The best FA recipe is $(z_{cut} = 0.6, \Delta z=0.06)$, which gives a largest $FoM = 4.6965$ (see Fig.~\ref{fig:1}).
This result holds true for all the five DE parameterizations, and thus is insensitive to a specific DE model (see table~\ref{tab:2}).

\item
(2) Flux-averaging JLA samples at $z_{cut} \geq 0.4$ will yield tighter DE constraints than the case without using FA; in contrast, the effects of varying $\Delta z$ are much smaller (see Fig.~\ref{fig:2}).

\item
(3) Using FA can significantly reduce the redshift-evolution of $\beta$ for the JLA SN sample (see Fig.~\ref{fig:3});
this implies that adopting FA technique can reduce the systematic uncertainties cased by $\beta$ evolution.

\item
(4) Adopting the best FA recipe will yield a larger fractional matter density $\Omega_{m}$ (see Fig.~\ref{fig:4})
and a universe with a smaller acceleration (see Fig.~\ref{fig:5}).
\end{itemize}

In summary, we present an alternative method of dealing with JLA data,
which can reduce the systematic uncertainties of SNe Ia and give the tighter DE constraints at the same time.
Our method will be useful in the use of SNe Ia data for precision cosmology.

There are a lot of related problems deserve further studies.
For examples, the best FA recipe obtained in this work can be used to constrain various DE models
\citep{Steinhardt1999,Caldwell2002,Li04,Wei05,Wang08b,Gao09,LLWZ09,Zhang2012,WLZL2016}.
Besides, it can be used to explore the dynamical evolution of DE EoS
by using various model-independent methods \citep{Huterer03,Huterer05,Huang09,WLL11,Li11,Gong13,hu2016,Mukherjee2016}.
In addition, based on the improved FA technique,
a series of related topics, such as dark sector interaction \citep{LLWWZ09}, massive neutrino \citep{LiYH2013},
cosmic age \citep{Wang08a,Lan10,WLL10}, cosmic fate \citep{Li12} and so on, need to be revisited.

In a recent work, Ma, Corasaniti and Bassett proposed a new Bayesian inference method to explore the JLA data,
by applying Bayesian graphs \citep{Bassett2016}.
Making use of this new analysis technique, they showed that the error bars of various model parameters can be significantly reduced.
It will be very interesting to explore the current SNe Ia data
by combining this Bayesian inference method with our best FA recipe.
This will be done in a future work.

\section*{Acknowledgments}

We are grateful to the referee for very helpful suggestions.
SW is supported by the National Natural Science Foundation of China under Grant No. 11405024
and the Fundamental Research Funds for the Central Universities under Grant No. 16lgpy50.
ML is supported by the National Natural Science Foundation of China (Grant No. 11275247, and Grant No. 11335012)
and a 985 grant at Sun Yat-Sen University.



\label{lastpage}

\end{document}